\documentstyle[twoside]{article}

\catcode`\@=11
\long\def\@makefntext#1{
\protect\noindent \hbox to 3.2pt {\hskip-.9pt  
$^{{\eightrm\@thefnmark}}$\hfil}#1\hfill}		%CAN BE USED 

\def\@makefnmark{\hbox to 0pt{$^{\@thefnmark}$\hss}}	%ORIGINAL 
	
\def\ps@myheadings{\let\@mkboth\@gobbletwo
\def\@oddhead{\hbox{}
\rightmark\hfil\eightrm\thepage}   
\def\@oddfoot{}\def\@evenhead{\eightrm\thepage\hfil
\leftmark\hbox{}}\def\@evenfoot{}
\def\sectionmark##1{}\def\subsectionmark##1{}}

\oddsidemargin=\evensidemargin
\newcounter{sectionc}\newcounter{subsectionc}\newcounter{subsubsectionc}
\renewcommand{\section}[1] {\vspace{12pt}\addtocounter{sectionc}{1} 
\setcounter{subsectionc}{0}\setcounter{subsubsectionc}{0}\noindent 
	{\tenbf\thesectionc. #1}\par\vspace{5pt}}
\renewcommand{\subsection}[1] {\vspace{12pt}\addtocounter{subsectionc}{1} 
	\setcounter{subsubsectionc}{0}\noindent 
	{\bf\thesectionc.\thesubsectionc. {\kern1pt \bfit #1}}\par\vspace{5pt}}
\renewcommand{\subsubsection}[1] {\vspace{12pt}\addtocounter{subsubsectionc}{1}
	\noindent{\tenrm\thesectionc.\thesubsectionc.\thesubsubsectionc.
	{\kern1pt \tenit #1}}\par\vspace{5pt}}

\newcommand{\textlineskip}{\baselineskip=13pt}

\def\abstracts#1#2#3{{
	\centering{\begin{minipage}{4.5in}\baselineskip=10pt\footnotesize
	\parindent=0pt #1\par 
	\parindent=15pt #2\par
	\parindent=15pt #3
	\end{minipage}}\par}} 

\renewenvironment{thebibliography}[1]
	{\frenchspacing
	 \ninerm\baselineskip=11pt
	 \begin{list}{\arabic{enumi}.}
        {\usecounter{enumi}\setlength{\parsep}{0pt}     
	 \setlength{\leftmargin 12.7pt}{\rightmargin 0pt} %FOR 1--9 ITEMS
         \setlength{\itemsep}{0pt} \settowidth
	{\labelwidth}{#1.}\sloppy}}{\end{list}}

\newcounter{itemlistc}
\newcounter{romanlistc}
\newcounter{alphlistc}
\newcounter{arabiclistc}

\def\@citex[#1]#2{\if@filesw\immediate\write\@auxout
	{\string\citation{#2}}\fi
\def\@citea{}\@cite{\@for\@citeb:=#2\do
	{\@citea\def\@citea{,}\@ifundefined
	{b@\@citeb}{{\bf ?}\@warning
	{Citation `\@citeb' on page \thepage \space undefined}}
	{\csname b@\@citeb\endcsname}}}{#1}}

\newif\if@cghi
\def\cite{\@cghitrue\@ifnextchar [{\@tempswatrue
	\@citex}{\@tempswafalse\@citex[]}}
\def\citelow{\@cghifalse\@ifnextchar [{\@tempswatrue
	\@citex}{\@tempswafalse\@citex[]}}
\def\@cite#1#2{{$\null^{#1}$\if@tempswa\typeout
	{IJCGA warning: optional citation argument 
	ignored: `#2'} \fi}}

\def\@refcitex[#1]#2{\if@filesw\immediate\write\@auxout
	{\string\citation{#2}}\fi
\def\@citea{}\@refcite{\@for\@citeb:=#2\do
	{\@citea\def\@citea{, }\@ifundefined
	{b@\@citeb}{{\bf ?}\@warning
	{Citation `\@citeb' on page \thepage \space undefined}}
	\hbox{\csname b@\@citeb\endcsname}}}{#1}}

\def\@refcite#1#2{{#1\if@tempswa\typeout
        {IJCGA warning: optional citation argument
	ignored: `#2'} \fi}}

\def\refcite{\@ifnextchar[{\@tempswatrue
	\@refcitex}{\@tempswafalse\@refcitex[]}}

\def\pmb#1{\setbox0=\hbox{#1}
	\kern-.025em\copy0\kern-\wd0
	\kern.05em\copy0\kern-\wd0
	\kern-.025em\raise.0433em\box0}

\def\fnt#1#2{\footnotetext{\kern-.3em
	{$^{\mbox{\scriptsize #1}}$}{#2}}}

\def\runninghead#1#2{\pagestyle{myheadings}
\markboth{{\protect\footnotesize\it{\quad #1}}\hfill}
{\hfill{\protect\footnotesize\it{#2\quad}}}}
\font\tenrm=cmr10
\font\tenit=cmti10 
\font\tenbf=cmbx10
\font\bfit=cmbxti10 at 10pt
\font\ninerm=cmr9

\font\eightrm=cmr8

\def\qed{\hbox{${\vcenter{\vbox{			%HOLLOW SQUARE
   \hrule height 0.4pt\hbox{\vrule width 0.4pt height 6pt
   \kern5pt\vrule width 0.4pt}\hrule height 0.4pt}}}$}}

\begin{document}

\runninghead{Rosu, ...
$\ldots$} {Rosu, q-intertwining between Schroedinger and Hermite
$\ldots$}

%Comment (HCR): produce frazele de mai sus la inceputul fiecarei pagini

\normalsize\textlineskip
\thispagestyle{empty}
\setcounter{page}{1}

%\copyrightheading{}                     %{Vol. 0, No.0 (1992) 000--000}

\vspace*{0.88truein}

%\fpage{1} %%%%%%%%%%%%%%%%%%%%%%%%%%%%%%%%%%%%%%%%%%%%%%%%%%%%%%%%%%%
\centerline{\bf BETWEEN SCHROEDINGER AND HERMITE:
SUPERSYMMETRIC PAIR      }
\centerline{\bf OF q-DEFORMED NON-LOCAL OPERATORS}
\vspace*{0.035truein}
%\centerline{\bf MANUSCRIPTS USING COMPUTER SOFTWARE\footnote{For
%the title, try not to use more than 3 lines. Typeset the title
%in 10 pt Times Roman, uppercase and boldface.}}
\vspace*{0.37truein}
\centerline{\footnotesize H.C. ROSU}
\vspace*{0.015truein}
\centerline{\footnotesize\it Instituto de F\'{\i}sica,
Universidad de Guanajuato, Apartado Postal E-143, 37150 Le\'on, Gto, Mexico}
%\baselineskip=10pt
%\centerline{\footnotesize\it City, State ZIP/Zone, Country}
\vspace*{0.225truein}
%\publisher{(March 8, 1999)}{(later or not necessary)}

\vspace*{0.21truein}
\abstracts{{\bf Abstract}.
A simple version of the q-deformed calculus is used to generate a pair of
q-nonlocal, second-order difference operators
by means of deformed counterparts of Darboux intertwining operators
for zero factorization energy.
These deformed non-local operators may be considered as
supersymmetric partners and their structure contains contributions originating
in both the Hermite operator and the quantum harmonic oscillator operator.
There are also extra $\pm x$ contributions.
The undeformed limit, in which all q-nonlocalities wash out, corresponds to
the usual supersymmetric pair of quantum mechanical harmonic oscillator
Hamiltonians.
The more general case of negative factorization energy is briefly
discussed as well.
}{}{}

%\vspace*{10pt}
%\keywords{The contents of the keywords}

\textlineskip                  %) USE THIS MEASUREMENT WHEN THERE IS
\vspace*{12pt}                 %) NO SECTION HEADING

\vspace*{1pt}\textlineskip	%) USE THIS MEASUREMENT WHEN THERE IS
%\section{General Appearance}    %) A SECTION HEADING
\vspace*{-0.5pt}
\noindent

%%%%%%%%%%%%%%%%%%%%%%%%%%%%%%%%%%%%%%%%%%%%%%
{\em PACS}: % 03.65.$\dagger$,
03.65.Ca, 03.65Fd


\bigskip

%------------------------------------------------------------------------------
%MACRO FOR COPYRIGHT BLOCK
%\def\eightcirc{
%\begin{picture}(0,0)
%\put(4.4,1.8){\circle{6.5}}
%\end{picture}}
%\def\eightcopyright{\eightcirc\kern2.7pt\hbox{\eightrm c}}

%\newcommand{\copyrightheading}[1]
%     {\vspace*{-2.5cm}\smalllineskip{\flushleft

  {\footnotesize {\bf To arXive or Not to arXive ? That's a good question !}}

  {\footnotesize except for the title and minor changes in the text,
   version of 3/8/99}

     {\footnotesize \copyright
        1999, by H.C. Rosu}
       %  }}

%------------------------------------------------------------------------------

\noindent
%%%%%%%%%%%%%%%%%%%%%%%%%%%%%%%%%%%%%%%%%%%%%%%%%%%%%%%%%%%%%%%%%%%%%

\newpage

%\pagebreak

%\textheight=7.8truein
%\setcounter{footnote}{0}
%\renewcommand{\thefootnote}{\alph{footnote}}

%\section{The Main Text}
\noindent

%{\bf 0}.
%\noindent
I present herein a simple $q$-deformed
procedure [\refcite{Jac}] for the basic case of the one-dimensional quantum
harmonic oscillator, by which I build a `supersymmetric' pair of q
nonlocal operators possessing terms whose $q\rightarrow 1$ limits
belong to either
Hermite polynomial operator or Schroedinger quantum oscillator operator.
There are $\pm x$ extra terms as well.
The procedure is based on the idea of using, as fundamental tools, a sort of
deformed counterparts of the intertwining operators encountered in the area of
Darboux transformations [\refcite{darb}].
I shall use their factorization property
to get the q-deformed second-order operators
which, being $q$ non-local, may be considered as more general than both the
usual Hermite one and the quantum mechanical harmonic oscillator operator.

The standard Hermite operator
${\hat{O}} _{H}$ reads ($D=d/dx$)
%%%%%%%%%%%%%%%
\begin{equation}
{\hat{O}}_{H}=D^2-2xD+2n~,
\end{equation}
%%%%%%%%%%%%%%%  -
and gives rise to the equation for the Hermite
polynomials $H_{n}(x)$, ${\hat{O}} _{H}H_{n}(x)= 0$.
Writing ${\hat{O}}_{H}=D^2-2xD-2+2(n+1)$, I shall treat $D^2-2xD-2$ as the
Fokker-Planck (FP) part of the Hermite operator for a stationary
transition-probability density, since $-2xD$ corresponds to
the $(\frac{dU}{dx})D$
drift contribution in the FP stationary operator (drift potential $U=-x^2$),
whereas $-2$ stands for the $d^2U/dx^2$ contribution of the FP drift. The
last term $2(n+1)$ gives the
departure of the Hermite operator from the corresponding FP stationary
operator for which polynomial oscillations are not allowed,
and in fact is responsible for turning the FP interpretation into a formal
one and not a physical one.
As well known for this basic case, by means of the functions $\phi _{n}=
e^{-x^2/2}H_{n}(x)$ one can go to the operator
%%%%%%%%%%%%%
\begin{equation}
{\hat{O}}_{\phi}=-D^2+[x^2- (2n+1)]~,
\end{equation}
%%%%%%%%%%%%%%%      2
which, in the $\phi _{n}$ space, is essentially the Schroedinger quantum
harmonic oscillator
operator up to a scaling, choose-of-units factor. One should notice that
this usage of the $\phi _{n}$ functions leads to the loss of
one half of the $\frac{d^2U}{dx^2}$ drift contribution. The remaining
half
%of the $U^{''}$ drift
gets the famous zero-point energy interpretation when the scaling
$\frac{1}{2}{\hat{O}}_{\phi}$ is performed.
In the FP interpretation, the latter scaling corresponds to setting the
diffusion constant equal to $1/2$ and provides the usual
quantum mechanical harmonic oscillator
wavefunctions $N_{n}\phi _{n}$, where $N_{n}=(2^n n!\sqrt{\pi})^{-1/2}$
is the normalization factor.

I now briefly recall that in the case of the one-dimensional Schroedinger
operator within the context of supersymmetric quantum mechanics (SUSYQM)
[\refcite{rsqm}]
the standard Darboux transformation operator reads
%%%%%%%%%%%%%
\begin{equation} \label{T}
{T}=-t_{u}(x)+D=-u^{\prime }(x)/u(x)+D~,
\end{equation}
%%%%%%%%%%%
where the prime denotes the derivative with respect to $x$.
When acting on the solutions $\psi _n(x)$ of the initial Schroedinger
equation
%%%%%%%%%%%%%%%%%%%%%%%%%%%%%%%%%%%
$
h_0\psi _n(x)=E_n\psi _n(x),
$
%%%%%%%%%%%%%%%%%%%%%%%%%%%
it transforms them into the solutions of another Schroedinger equation
$h_1\varphi _n(x)=E_n\varphi _n(x)$,
%%%%%%%%%%%%%
$\varphi _n(x)=N_n{T}\psi _n(x)$,
%6bis
%%%%%%%%%%%%%%
with the same eigenvalues $E_n$. Henceforth, I will put the
ground state energy equal
to zero, $E_0=0$, since this does not affect in any way the results.
The new exactly solvable Hamiltonian has the form
$h_1=h_0+\Delta V(x)$, where the potential difference is of Darboux type
$\Delta V(x)=-2(\ln
u)^{\prime \prime }$. The function $u=u(x)$
is a so-called transformation function,
being a solution of the initial Schroedinger equation
%%%%%%%%%%%%%%%%
$
h_0u(x)=\epsilon u(x),
$
%%%%%%%%%%%%%%%%%
with $\epsilon \leq 0$ usually known as the factorization energy.
It is well established that
when $\epsilon <0$ one can work with a nodeless transformation function by
performing an analytic continuation [\refcite{bsam}]. Thus,
$u(x)\ne 0$ for any value of the variable %$\forall x\in(0,\infty )$
and $1/u(x)$ is not a square integrable function.
In this  case $u\notin {\cal H}_{1}$ and the
set $\{\mid \varphi _n\rangle \}$
is a complete basis in the Hilbert space ${\cal H}_{1}$ provided
the initial system $\{\mid \psi _n\rangle \}$ is complete.
%In terms of the supersymmetric quantum mechanics this case corresponds to a
%broken supersymmetry.
%\noindent
The operator ${T}^{+}=-t_{u}(x)-D$ provides the backward transformation
%%%%%%%%%%%%%%%%%%%%%
$
\label{psin}|\psi _n\rangle =N_n {T}^{+}|\varphi _n\rangle ,
$
%%%%%%%%%%%%%%%%%%%%%
and together with ${T}$ allows for the following factorizations
%%%%%%%%%%%
\begin{equation}
{T}^{+}{T}=h_0-\epsilon ,\quad {T}{T}^{+}=h_1-\epsilon \ .
%\eqno(15)
\end{equation}
%%%%%%%%%%%
The operators ${T}$ and ${T}^{+}$ are
well defined $\forall \psi \in {\cal H}_{1}$ and are conjugated to each
other with respect to the inner product in the ${\cal H}_{1}$ space.

My purpose now is to get $q$-deformed second-order
operators by means of deformed counterparts of the
aforementioned intertwining operators. I still have to present some
definitions and rules of the deformed calculus.
%%%%%%%%%%%%
Since the independent variable is maintained commutative, the employed
version of the deformed calculus is similar to that previously used
by some authors to deform the Coulomb problem [\refcite{coul}].
Symmetric definitions of the
q-number $[x]_q=\frac{q^{x}-q^{-x}}{q-q^{-1}}$
and $q$-derivative
%%%%%%
\begin{equation}
D_{q}f(x)=\frac{f(qx)-f(q^{-1}x)}{x(q-q^{-1})}~
\end{equation}
%%%%%%%%%%%    3
are used together with
some basic rules of Jackson's calculus [\refcite{Jac}] such as
%%%%%%%
$D_{q} x^{n}=[n]_{q}x^{n-1}$, $D_{q}^{2}x^n=[n]_q[n-1]_qx^{n-2}$,
$D_{q}(FG)=(D_{q}F)G(qx)+F(q^{-1}x)(D_{q}G)$ for any two functions $F$ and $G$,
respectively.
%%%%%%
%are of use in the calculations.
The definition of the $q$-exponential is
%%%%%%%
\begin{equation}
e_{q}(x)= \sum _{n=0}^{\infty}\frac{x^{n}}{[n]_{q}!}~,
\end{equation}
%%%%%%%%%%%%%%  4
which reduces to the usual exponential function as $q\rightarrow 1$, and
moreover is invariant under $q\rightarrow q^{-1}$.

%\noindent
The main idea of this work is based on the following scheme. First,
to employ as Darboux transformation functions
deformed counterparts of the oscillator vacua
$\psi_{q}\propto e_q(\beta x^2)$,
where $\beta =\pm 1/2$ for the irregular and regular vacuum, respectively.
Second, to exploit the factorization property of
first-order deformed operators of the form
%%%%%%%%%%%%%%%
\begin{equation}
T_{+}^{q}=D_{q}-\frac{D_{q}\psi_{q}}{\psi_{q}}=
D_q % - \frac{[\gamma]_{q}}{x}\epsilon _{q}(x^2)
-\beta _{q}(x^2) x~,
%\eqno(3.8)
\end{equation}
%%%%%%%%%%%%%%%%  7
\begin{equation}
T_{-}^{q}=-D_{q}-\frac{D_{q}\psi_{q}}{\psi_{q}}=
-D_q%-\frac{[\gamma]_{q}}{x}\epsilon _{q}(x^2)
-\beta _{q}(x^2)x~,
%\eqno(3.8)
\end{equation}
%%%%%%%%%%%%%%%%  8
where
%%%%%%%%%%%%%%%
\begin{equation}
\beta _{q}(x^2)=\beta %q^{-\gamma}
\left(\frac{qe_{q}(q\beta x^{2})+
q^{-1}e_{q}(q^{-1}\beta x^{2})}{e_{q}(\beta x^2)}\right)~.
\end{equation}
%%%%%%%%%%%%%%%%%  9
The form of $\beta _{q}(x^2)$ is a result of Jackson's calculus rules.
As one can see, the $T_{+}^{q}$ and $T_{-}^{q}$ operators have been written
by analogy to the continuous intertwining operators.
A straightforward calculation gives the second-order deformed
operators that can be obtained
from the products $T_{-}^{q}T_{+}^{q}$ and
$T_{+}^{q}T_{-}^{q}$, respectively. One gets
%%%%%%%%%%%%%%%%%%%%%%%
\begin{equation}
{\hat{O}}_{b}^{q}\equiv T_{-}^{q}T_{+}^{q}=
-D_{q}^{2}-
%\Bigg[
%\left(
[(\Delta \beta _{q})xD_{q}]%\Bigg]+
+[
\beta _{q}^{2}(x^2)x^2
]_{\rightarrow x}
+[q(D_{q}\beta _{q}(x^2))x]_{\rightarrow qx}%\Bigg]
+[\beta _{q}(q^{-2}x^2)]_{\rightarrow qx}%\Bigg]
\end{equation}
%%%%%%%%%%%%%%%%%%%%%%%%%%   10
and
\begin{equation}
{\hat{O}} _{f}^{q}\equiv T_{+}^{q}T_{-}^{q}=
-D_{q}^{2}+
[(\Delta \beta _{q})x%\right)
D_{q}]%\Bigg]+
+[\beta _{q}^{2}(x^2)x^2]
_{\rightarrow x}
-[q(D_{q}\beta _{q}(x^2))x]_{\rightarrow qx}%\Bigg]
-[\beta _{q}(q^{-2}x^2)]_{\rightarrow qx} %\Bigg]~,
\end{equation}
%%%%%%%%%%%%%%%%%%%%%%%%%%     19
where
%%%%%%%%%%%%%%%%%%
\begin{equation}
\Delta \beta _{q}=\beta _{q}(x^2)- q^{-1}\beta _{q}(q^{-2}x^2)~.
\end{equation}
%%%%%%%%%%%%%%%%%%%%%%%%%%

\noindent
The operators ${\hat{O}} _{b}^{q}$
and ${\hat{O}} _{f}^{q}$ may be considered
as supersymmetric partners since they have been built according to
the well-known SUSYQM method. At this point
one should notice the interesting mixed structure of the two
$q$ non-local operators that entail parts of both
${\hat{O}}_{H}$ and ${\hat{O}}_{\phi}$.
%and a ``spatially-spreaded" zero point part.
The directional subindices indicate the argument of the
solution on which the nonoperatorial parts act. The two operators are
nonlocal operators whose space of solutions are the
functions $\phi ^{(q)}_{n}(x)\propto e_{q}(-x^2/2)H^{(q)}_{n}(x)$, where
the deformed Hermite polynomials can be defined by a $q$ deformed
Rodrigues representation
%%%%%%%%%%%
\begin{equation}
H^{(q)}_{n}=(-1)^{n}e_{q}(x^2)D_{q}^{n}(e_{q}(-x^2))~.
\end{equation}
%%%%%%%%%%%%%%%%%%
%The main feature of the operators ${\hat{O}}_{b}^{q}$
%and ${\hat{O}}_{f}^{q}$ is the $q$-nonlocality.
Notice that the $D_{q}$ terms in (10) and (11)
are identical but opposite in sign
and correspond to the first derivative drift term in the
Hermite differential operator. Of course, if one prefers the FP interpretation
the two operators should be multiplied by (-1). Writing the finite difference
$x(q-q^{-1})=x\Delta q=\Delta _{q} x$, which
for $q\rightarrow 1$ is assumed to be a $q$ scaling way of going to the
infinitesimal limit $dx$, the $q$ drift parts go to zero, whereas in the
same limit the potential and zero point sectors take forms
identical to those of the undeformed case. More precisely,
the undeformed limits read
%%%%%%%%%%%%%%%%%
\begin{equation}
{\hat{O}} _{b}^{1}\equiv h_{0}=-D^2 %+\frac{\gamma(\gamma -1)}{x^2}+
+\beta _{1}^{2}x^2+\beta _{1}%(2\gamma+1)
\end{equation}
%%%%%%%%%%%%%%%%  14
and
\begin{equation}
{\hat{O}} _{f}^{1}\equiv h_{1}=-D^2 %+\frac{\gamma(\gamma +1)}{x^2}+
+\beta _{1}^{2}x^2-\beta _{1}~.%(2\gamma-1)~.
\end{equation}
%%%%%%%%%%%%%%%%    15
(14) and (15) are the usual quantum mechanical
supersymmetric partner Hamiltonians for this case.

%%%%%%%%%%%%%%%%%%%%%%%%%%%%%%%%%%%%%%%%%%%%%%%%%%%%%%%%%%%%%%%
%%%%%%%%%%%%%%%%%%%%%%%%%%%%%%%%%%%%%%%%%%%%%%%%%%%%%%%%%%%%%%%
In SUSYQM terminology, only the case of zero factorization energy
$\epsilon =0$ has been tackled up to now, but following a suggestion of
Bagrov and Samsonov [\refcite{bsam}], there is no difficulty to sketch the
procedure for the more general case $\epsilon <0$.
First, the deformed Schroedinger solution corresponding to the excited
harmonic oscillator states can be sought in the form
%%%%%%%%%%
\begin{equation}
\psi _{n}^{(q)}(x)\propto H_{n}^{(q)}(x)e_{q}(-x^2/2)~.
\end{equation}
%%%%%%%%%%%%%%%  16

%%%%%%%%%%%%%%%%%%%%%%%%%%%%%%%%%%%%%%%%%%%%%%%%%%%%%%%%%%%%%%%%%%%%%%
\noindent
Next, in order to avoid any singularities,
it is convenient to perform an $i$-rotation
$x\rightarrow ix$, leading to
%%%%%%%%%%%%%%%%
\begin{equation}
u_{p}^{(q)}(x)\propto H_{p}^{(q)}\left( ix\right)
e _{q} \left( x^2/2\right) ,%\quad y=-x^2/2 ,
\quad p=0,1,2,3\ldots
\end{equation}
%%%%%%%%%%%%%%   17
The undeformed functions $u_{p}^{(1)}$ are solutions of
$h_{0}u_{p}^{(1)}=-(p+1)u_{p}^{(1)}$ and are nodeless on the full line for
even $p=2k$. Therefore, they have been used by Bagrov and Samsonov as
Darboux transformation functions to generate a family of regular potentials,
which, according to an interpretation due to
Veselov and Shabat [\refcite{vs}], has a spectrum made up
of $2k+1$ segments with equidistant levels.
This immediately suggests
using $u_{p}^{(q)}(x)$ for even $p=2k$ as Darboux transformation
functions in the deformed case.
%since
%$H_{2k}^{(q)}(ix)\ne 0$ on the entire axis
%$\forall x\ne 0$ when $y=-x^2/2$$(<0)$
%and hence, the function $u_{2m}^{(q)}(x)$ is nodeless.
%%%%%%%%%%%%%%%%%%%%%%%%%%%%%%%%%%%%%%%%%%%%%%%%
Thus, the intertwining operators can be calculated according to
$T_{2k,+}^{q}=D_{q}-\frac{D_q u_{2k}^{(q)}}{u_{2k}^{(q)}}$ and
$T_{2k,-}^{q}=-D_{q}-\frac{D_q u_{2k}^{(q)}}{u_{2k}^{(q)}}$, and again by
exploiting the
factorization property one is led to second-order deformed, non-local
operators of more complicated formulas than (10) and (11) for which they
are not written down here.

%\vspace{3mm}

%{\bf 3}.
In conclusion, a pair of $q$ non-local
second-order $q$-differential ($q$-difference) operators have been introduced
in this work by means of a particular $q$-deformed
intertwining based on $q$-deformed oscillator vacua as Darboux transformation
functions.
These operators display a mixed structure between the Hermite operator,
to which they are similar as regards the first derivative term, and the
quantum mechanical oscillator operator, to which they are similar as regards
the $x^2$ potential and zero-energy contributions.
%the latter being $q$ non-local.
On the other hand, they present
a supplementary $q$ non-local $\pm x$ potential contribution with no
counterpart in either Hermite polynomial operator or
Schroedinger $x^2$ oscillator operator. All these features suggest many
possible applications, e.g., in mesoscopic physics. A more general case
corresponding
to negative factorization energies of the type $\epsilon _{m}=-(m+1)$,
where $m$ is an even positive integer, has also been briefly described.

\vspace{0.5cm}
\noindent
This work has been supported in part by CONACYT project 458100-5-25844E .

\vspace{3mm}

\end{document}